\documentclass[prb,twocolumn]{revtex4-2}

\usepackage{graphicx}
\usepackage{dcolumn}
\usepackage{bm}
\usepackage{mathrsfs}
\usepackage{subfigure}
\usepackage{natbib}
\usepackage{amsmath}
\usepackage{amssymb}
\usepackage{Jonasmacros}
\usepackage{pifont}
\usepackage{times}
\usepackage{txfonts}

\usepackage[colorlinks,citecolor=RedOrange,linkcolor=Fuchsia]{hyperref}
\usepackage[usenames,dvipsnames,svgnames]{xcolor}

\date{\today} 
\begin{document}

\title{An argument why the Spinterface model cannot explain the chirality induced spin selectivity effect}

\author{J. Fransson}
\email{Jonas.Fransson@physics.uu.se}
\affiliation{Department of Physics and Astronomy, Box 516, 752 37, Uppsala University, Uppsala, Sweden}




\begin{abstract}
In the context of chirality induced spin selectivity effect, it has been argued that a chiral molecule when adsorbed on a metal facilitates the formation of a local spin moment at the interface between the metal and molecule, given a strong spin-orbit coupling in the metal. The possibility for such spin moment formation is analyzed in terms of general arguments and effective modeling of a pertinent set-up. The conclusion from this analysis is that a strong spin-orbit coupling in the metal does not provide a sufficient mechanism to sustain a stabilized spin moment at the interface. It is, moreover, shown that an electron flux in to or out from the molecule does not provide conditions for a spin moment formation, regardless of whether the flux is spin-polarized or not.
\end{abstract}
\maketitle

\section{Introduction}
To some degree, magnetism and magnetization dynamics can be captured within the framework of the  Landau-Lifshitz-Gilbert equation \cite{LandauLifshitzEq,IEEETransMagn.40.3443}. The main restrictions have to do with that the equation applies to magnetic quantities that can be addressed as classical variables. Hence, ferromagnetic domain structures and localized spin moments represent typical contexts where the Landau-Lifshitz-Gilbert equation can be successfully utilized and yielding trustworthy results. A technical reason for these limitations is that it merely addresses transversal spin variations of a vector quantity that does not vary in length, see for instance Ref. \citenum{PhysRevLett.108.057204}. Hence, one can conveniently study the dynamics of spin waves or an emergent collective spin moment from an ensemble of classical vector spin that may rotate freely in space.

By contrast, the Landau-Lifshitz-Gilbert equation is not applicable to study the emergence of a spin density from the imbalance between the electronic spin-degrees of freedom. Such quantum phenomenology is beyond the classical nature of the Landau-Lifshitz-Gilbert equation.

A spin phenomenon which has been debated whether it is caused by a local spin density emerging at the interface between chiral molecules and a metal is the \emph{chirality induced spin selectivity} effect. It has been theoretically demonstrated that a chiral molecule by electron correlations acquires a non-trivial enantioselective spin-distribution when interfaced with a metal \cite{JPhysChemLett.16.4346,arXiv:2509.17817}. Moreover, the spin-distribution arises regardless of whether the metal sustains a spin-orbit coupling or non-trivial spin-texture, e.g, ferromagnetism or anti-ferromagnetism. An alternative proposal, referred to as the \emph{spinterface} theory \cite{JACS.143.14235,JACS.146.32795,ACSNano.19.37484}, builds on the presumption that a local magnetization density emerges at or around the interface between the metal and the molecule, initiated by the molecular chirality, however, entirely driven by the metallic spin-orbit coupling. The basic assumptions are that (i) moving electrons generate a Biot-Savart magnetic field which, according to the Faraday-Lens law, induces a small magnetization density in the metal at the interface, and (ii) the induced magnetization density transiently accumulates into a localized spin-moment and becomes strongly directed along the transport direction defined by the junction. The second point is, in the spinterface theory, justified by the Landau-Lifshitz-Gilbert equation in which a local exchange field and damping are generated by interactions with the surrounding charge density.

The current article is a polemic interjection to why the spinterface \cite{JACS.143.14235,JACS.146.32795,ACSNano.19.37484} theory, while phenomenologically interesting, cannot be based on the two of arguments quoted above. Whereas the impassible application of the Landau-Lifshitz-Gilbert equation has to do with the dichotomy between quantum versus classical quantities, the question whether a spin-moment large enough to be considered as classical may be generated at the interface can be straight-forwardly investigated in terms of a simple modeling.

Before going into the details of the modeling and calculations, it is interesting to note that Au have been reported to acquire magnetic responses in nanocluster and -particles \cite{PhysLettA.263.406,PhysRevLett.93.116801,JNanopartRes.12.177,GoldBull.44.3,Small.10.907,ACSOmega.2.2607,NatPhotonics.14.365,Nanoscale.12.19797,JPhysChemC.125.20482} and in thin films \cite{ApplPhysLett.88.222502,ACSNano.7.6691}. These observations are relevant since Au is a used as a substrate for the molecules in a vast majority of the experiments on the chirality induced spin selectivity effect, see for example Refs.  \citenum{Science.283.814,Science.331.894,NanoLett.11.4652,JACS.138.15551,ACSNano.11.7516,NatComms.8.14567,AdvMater.31.1904206,ACSNano.14.15983,AngewChemie.59.14671,JACS.142.17572,ACSAppliedMaterInterfaces.14.30813,NatComms.13.3356,NanoLett.23.8280,AdvMater.36.2406347,AngewChemie.63.202315146}.

Magnetic moments have been observed in pristine Au nanoclusters \cite{JNanopartRes.12.177}, core-shell structure \cite{GoldBull.44.3}, nanoparticles decorated with various ligand molecules \cite{PhysRevLett.93.116801,Small.10.907,ACSOmega.2.2607,Nanoscale.12.19797,JPhysChemC.125.20482}, and in optically excited nanoparticles \cite{NatPhotonics.14.365}. In all but one of these examples, stable ferromagnetic moments were only observed at temperatures far below room temperature. In icosahedral gold nanoparticles \cite{JNanopartRes.12.177}, a spontaneous magnetic moment was concluded to persist beyond room temperature. Nevertheless, the coercive field associate  with the ferromagnetism in all nanoparticles and -cluster is typically less than 1,000 \O\  ($\sim 0.1$ K). Ferromagnetism was measured in Au thin films at low temperatures, with a coercive field less than 100 \O, whereas the ferromagnetic signature was either completely lost \cite{ApplPhysLett.88.222502} or very weak \cite{ACSNano.7.6691} t room temperature.

Despite the factual status that ferromagnetism and magnetic moments have been observed in different types of Au compositions, there is no doubt that these observations have little bearing on the experimental reality pertaining to measurements of the chirality induced spin selectivity effect. Especially since most of the pertinent reported experiments  have been performed at room temperature. Certainly, decorating a several nm thick Au surface with chiral molecules, may give rise to localized magnetic moments in Au in the proximity of the metal-molecule interface. However, without further knowledge about the actual conditions that prevails in the experimental set-ups, an \emph{ad hoc} inclusion of a finite magnetic moment does not explain the origin of the chirality induced spin selectivity effect. To this end, it is necessary to theoretically justify whether a localized magnetic moment is dependable in this context.

The main question concerning the origin of a possible stabilization of a magnetic moment at the interface between the metal and the chiral molecule has to do with whether the emergent magnetization can be considered in a local moment picture or as an itinerant electron spin-polarization. For noble metals as well as relevant organic molecules, there are no unpaired $d$-electrons which could constitute a localized spin moment. Moreover, while there in principle could be a charge transfer between the molecule and the metal which could unleash a paired $d$-electron state and, subsequently enable a local spin moment, there is nothing obvious in the set-up that should lead to such sequence of event. Therefore, the localized moment picture would have to be dismissed until proofs or argument of the opposite can be presented. The conclusion is, hence, that any emergent spin moment has to be associated with  the current carrying itinerant electrons.

\section{Electrons in the metal}
The question to be addressed here pertains to the governing mechanisms for a localized magnetic moment induced in a metal by the presence of a defect. The defect is assumed to possess internal degrees of freedom, such as, electron-vibron and/or Coulomb interactions. It is, moreover, assumed that there is a particle exchange between the metal and the molecule at the interface. Raised by the assumptions forming the basis for the spinterface model \cite{JACS.143.14235,JACS.146.32795,ACSNano.19.37484}, it is particularly interesting to elucidate whether the metallic spin-orbit coupling is a mechanism that sustains the accumulation of a local spin density, and whether this spin-density can be responsible for the chirality induced spin selectivity effect.

A concrete yet reasonably simple model of the composite system comprising a molecule adsorbed onto the surface is formulated using the Hamiltonian
\begin{align}
\Hamil=&
	\sum_\bfk\psi^\dagger_\bfk\bfepsilon_\bfk\psi_\bfk
	+
	\Hamil_\text{mol}
	+
	\int
		\Bigl(\psi^\dagger(\bfr)\bfv(\bfr)\psi_1+H.c.\Bigr)
	d\bfr
	.
\label{eq-Hamiltonian}
\end{align}
In this model, the $2\times2$-matrix $\bfepsilon_\bfk=\dote{0}(\bfk)\sigma^0+\bfepsilon_1(\bfk)\cdot\bfsigma$ define the band structure of the electrons in the metal and $\bfv(\bfr)=v_0(\bfr)\sigma^0+\bfv_1(\bfr)\cdot\bfsigma$ denotes hybridization matrix element between states in the metal and molecule. Here, $\sigma^0$ and $\bfsigma$ denote the $2\times2$-identity matrix and vector of Pauli matrices, respectively. The electron destruction (creation) spinors $\psi_\bfk=\int\psi(\bfr)e^{-i\bfk\cdot\bfr}d\bfr$ ($\psi^\dagger_\bfk=\int\psi^\dagger(\bfr)e^{i\bfk\cdot\bfr}d\bfr$) and $\psi_1$ ($\psi_1^\dagger$) operate in the metal and molecule, respectively. In this notation, the hybridization term in the Hamiltonian can be written $\sum_\bfk\psi^\dagger_\bfk\bfv_\bfk\psi_1+H.c.$, where $\bfv_\bfk=\int\bfv(\bfr)e^{-i\bfk\cdot\bfr}d\bfr$.

An obvious weakness with this model is the lack of particle interactions within the metal. The justification for an independent particle description is, nonetheless, the massive Coulomb screening between the electrons, generating effectively non-interacting quasi-particles. Such conditions pertains to itinerant electrons, which in the noble metals comprise the $s$- and $p$-bands. Correlated electrons in the $d$-band can be omitted since this band is typically located far below the Fermi level and are, hence, filled and most likely inactive under the current conditions. This simplification allows analytical calculations and deductions of the basic mechanisms. The conclusions drawn here can be further analyzed using, e.g., the homogeneous electron gas in which charge density-density correlations are included. Nevertheless, an explicit presence of the Coulomb repulsion among the electrons in the metal is not necessary in the formal discussion below since the electronic properties are expressed in terms of single-electron Green functions. The expressions formulated in terms of the Green functions are not changed upon inclusion of the Coulomb interaction and, hence, the initial discussion below is general.


\subsection{Itinerant spin moment}
The time-dependent electron spin moment $\av{\bfs(\bfr,t)}$ for an electron at some coordinate $\bfr$ in the metal can be accessed through the single electron Green function $\bfG_{\bfk\bfk'}(t,t')=\eqgr{\psi_\bfk(t)}{\psi^\dagger_{\bfk'}(t')}$ thanks to the relation
\begin{align}
\av{\bfs(\bfr,t)}=&
	(-i)\frac{1}{2}
	\lim_{t'\rightarrow t^+}{\rm sp}\bfsigma
		\int
			\bfG^<_{\bfk\bfk'}(t,t')e^{i(\bfk-\bfk')\cdot\bfr}
		\frac{d\bfk}{\Omega}
		\frac{d\bfk'}{\Omega}
	.
\end{align}
Here, $\bfG_{\bfk\bfk'}^<(t,t')$ denotes the lesser form of $\bfG_{\bfk\bfk'}(t,t')$ and represents the density of occupied electron states. The physics associated with the presence of the defect $\Hamil_\text{mol}$ can be captured as a self-energy $\bfSigma_{\bfk\bfk'}(t,t')$ in the Dyson equation
\begin{align}
\bfG_{\bfk\bfk'}(t,t')=&
	\bfg_{\bfk\bfk'}(t,t')
\label{eq-Dyson}
\\&
	+
	\int
		\bfg_{\bfk\bfp}(t,\tau)
		\bfSigma_{\bfp\bfq}(\tau,\tau')
		\bfG_{\bfq\bfk'}(\tau',t')
	\frac{d\bfp}{\Omega}
	\frac{d\bfq}{\Omega}
	d\tau
	d\tau'
	,
\nonumber
\end{align}
where $\bfg_{\bfk\bfk'}(t,t')$ denotes the unperturbed Green function. Therefore, the lesser Green function can be obtained using the relation
\begin{align}
\bfG^<_{\bfk\bfk'}(t,t')=&
	\int
		\bfG_{\bfk\bfp}^r(t,\tau)
		\bfSigma_{\bfp\bfq}^<(\tau,\tau')
		\bfG_{\bfq\bfk'}^a(\tau,t')
	\frac{d\bfp}{\Omega}
	\frac{d\bfq}{\Omega}
	d\tau
	d\tau'
	,
\end{align}
where $\bfG_{\bfk\bfk'}^{r/a}$ denotes the retarded/advanced form of $\bfG$.

By the general expansion for a $2\times2$-matrix, $\bfG=G^0\sigma^0+\bfG^1\cdot\bfsigma$, the electronic properties are conveniently partitioned into its charge and spin properties by setting $G^0={\rm sp}\bfG/2$ and $\bfG^1={\rm sp}\bfsigma\bfG/2$, respectively. The trace over spin 1/2 space opens up for the natural partitioning of the induced spin-moment as
\begin{align}
\av{\bfs(\bfr,t)}=&
	\av{\bfs(\bfr,t)}_\text{mol}
	+
	\av{\bfs(\bfr,t)}_\text{sub}
	+
	\av{\bfs(\bfr,t)}_\text{mix}
	.
\end{align}
In this form, the subscripts refer to the origins of the sources for spin-polarization of the electron. The three components can be defined as
\begin{widetext}
\begin{subequations}
\label{eq-s}
\begin{align}
\av{\bfs(\bfr,t)}_\text{mol}=&
	(-i)
	\int
		G^{0,r}_{\bfk\bfp}(t,\tau)
		\bfSigma^{1,<}_{\bfp\bfq}(\tau,\tau')
		G^{0,a}_{\bfq\bfk}(\tau',t)
		e^{-i(\bfk-\bfk')\cdot\bfr-i(\bfp-\bfq)\cdot\bfr_0}
	\frac{d\bfp}{\Omega}
	\frac{d\bfq}{\Omega}
	\frac{d\bfk}{\Omega}
	\frac{d\bfk'}{\Omega}
	d\tau
	dt\tau'
	,
\label{eq-smol}
\\
\av{\bfs(\bfr,t)}_\text{sub}=&
	(-i)
	\int
		\biggl\{
			G^{0,r}_{\bfk\bfp}(t,\tau)
			\Sigma^{0,<}_{\bfp\bfq}(\tau,\tau')
			\bfG^{1,a}_{\bfq\bfk}(\tau',t)
			+
			\bfG^{1,r}_{\bfk\bfp}(t,\tau)
			\Sigma^{0,<}_{\bfp\bfq}(\tau,\tau')
			G^{0,a}_{\bfq\bfk}(\tau',t)
\nonumber\\&
			+
			i\Sigma^{0,<}_{\bfp\bfq}(\tau,\tau')
			\bfG^{1,r}_{\bfk\bfp}(t,\tau)\times\bfG^{1,a}_{\bfq\bfk}(\tau',t)
		\biggr\}
		e^{-i(\bfk-\bfk')\cdot\bfr-i(\bfp-\bfq)\cdot\bfr_0}
	\frac{d\bfp}{\Omega}
	\frac{d\bfq}{\Omega}
	\frac{d\bfk}{\Omega}
	\frac{d\bfk'}{\Omega}
	d\tau
	dt\tau'
	,
\label{eq-ssub}
\\
\av{\bfs(\bfr,t)}_\text{mix}=&
	(-i)
	\int
		\biggl\{
			\Bigl(
				\bfG^{1,r}_{\bfk\bfp}(t,\tau)
				\cdot
				\bfSigma^{1,<}_{\bfp\bfq}(\tau,\tau')
			\Bigr)
			\bfG^{1,a}_{\bfq\bfk}(\tau',t)
			+
			\bfG^{1,r}_{\bfk\bfp}(t,\tau)
			\Bigl(
				\bfSigma^{1,<}_{\bfp\bfq}(\tau,\tau')
				\cdot
				\bfG^{1,a}_{\bfq\bfk}(\tau',t)
			\Bigr)
			-
			\Bigl(
				\bfG^{1,r}_{\bfk\bfp}(t,\tau)
				\cdot
				\bfG^{1,a}_{\bfq\bfk}(\tau',t)
			\Bigr)
			\bfSigma^{1,<}_{\bfp\bfq}(\tau,\tau')
\nonumber\\&
			+
			i
			\bfG^{1,r}_{\bfk\bfp}(t,\tau)
				\times
			\bfSigma^{1,<}_{\bfp\bfq}(\tau,\tau')
			G^{0,a}_{\bfq\bfk}(\tau',t)
			+
			i
			G^{0,r}_{\bfk\bfp}(t,\tau)
			\bfSigma^{1,<}_{\bfp\bfq}(\tau,\tau')
				\times
			\bfG^{1,a}_{\bfq\bfk}(\tau',t)
		\biggr\}
		e^{-i(\bfk-\bfk')\cdot\bfr-i(\bfp-\bfq)\cdot\bfr_0}
	\frac{d\bfp}{\Omega}
	\frac{d\bfq}{\Omega}
	\frac{d\bfk}{\Omega}
	\frac{d\bfk'}{\Omega}
	d\tau
	dt\tau'
\label{eq-smix}
	.
\end{align}
\end{subequations}
\end{widetext}
This partitioning is organized by using the fact that the self-energy that arises due to the presence of the molecule, can also be partitioned according to $\bfSigma_{\bfp\bfq}^<=\Sigma_{\bfp\bfq}^{0,<}\sigma^0+\bfSigma_{\bfp\bfq}^{1,<}\cdot\bfsigma$.

The spin-moment components as detailed in Eq. \eqref{eq-s} can be further analyzed by reducing the Green functions $G_{\bfk\bfk'}^{0,r/a}$ and $\bfG^{1,r/a}$ to their corresponding unperturbed forms on which the defect has no influence. Then, the spin-dependence is solely originating from the molecule ($\bfSigma^{1,<}$) in the component $\av{\bfs(\bfr,t)}_\text{mol}$ whereas it is a property of the metal ($\bfG^{1,r/a}$) in $\av{\bfs(\bfr,t)}_\text{sub}$. The spin-moment accounted for in $\av{\bfs(\bfr,t)}_\text{mix}$ cannot be separated in this sense as it depends on the spin-polarization in both the molecule ($\bfSigma^{1,<}$) and the metal ($\bfG^{1,r/a}$).

A direct analysis suggests two conclusions. First, Eq. \eqref{eq-smol} indicates that the spin-dependence in the molecule may be transferred to the substrate also in the absence of a spin-texture in the metal. Second, from Eqs. \eqref{eq-ssub} and \eqref{eq-smix} it is clear that the existence of a spin-texture in the metal may be locally modified in the presence of a molecule. Moreover, this modification occurs regardless of whether there is a molecular spin-polarization, Eq. \eqref{eq-smix}, or not, Eq. \eqref{eq-ssub}.

\subsection{Self-energy}
The model introduced in Eq. \eqref{eq-Hamiltonian} leads to the self-energy $\bfSigma^<_{\bfp\bfq}(t,t')=\delta(\bfp-\bfq)\bfv_\bfp\bfG_1^<(t,t')\bfv^\dagger_\bfp$, where $\bfG_1^<$ is the lesser Green function for electrons in the molecule. The assumption is that the particle exchange between the metal and molecule occurs at one molecular site only, an assumption which can be relaxed at the cost of further complexity of the modeling. The self-energy components are formally defined by
\begin{subequations}
\label{eq-Sigma}
\begin{align}
\Sigma^{0,<}_{\bfp\bfq}=&
	\Bigl(
		v_{0\bfp}v_{0\bfq}^*
		+
		\bfv_{1\bfp}\cdot\bfv_{1\bfq}^*
	\Bigr)
	G_1^{0,<}
\label{eq-Sigma0}\\&
	+
	\bfv_{1\bfp}\cdot\bfG_1^<v_{0\bfq}^*
	+
	v_{0\bfp}\bfv_{1\bfq}^*\cdot\bfG_1^{1,<}
	-
	i\Bigl(
		\bfv_{1\bfp}\times\bfv_{1\bfq}^*
	\Bigr)
	\cdot\bfG_1^{1,<}
	,
\nonumber\\
\bfSigma^{1,<}_{\bfp\bfq}=&
	\Bigl(
		v_{0\bfp}\bfv_{1\bfq}^*
		+
		\bfv_{1\bfp}v_{0\bfq}^*
	\Bigr)G_1^{0,<}
	+
	\Bigl(
		v_{0\bfp}v_{0\bfq}^*
		-
		\bfv_{1\bfp}\cdot\bfv_{1\bfq}^*
	\Bigr)\bfG_1^{1,<}
\nonumber\\&
	+
	\bfv_{1\bfp}\cdot\bfG_1^{1,<}\bfv_{1\bfq}^*
	+
	\bfv_{1\bfp}\bfv_{1\bfq}^*\cdot\bfG_1^{1,<}
\label{eq-Sigma0}\\&
	+
	i\bfv_{1\bfp}\times\bfv_{1\bfq}^*G_1^{0,<}
	+
	i\bfv_{1\bfp}\times\bfG_1^{1,<}v_{0\bfq}^*
	+
	iv_{0\bfp}\bfv_{1\bfq}^*\times\bfG_1^{1,<}
	.
\nonumber
\end{align}
\end{subequations}

The complicated structure of the self-energy is due to the possibility that the hybridization between the metallic and molecular states, captured in the hybrization potential $\bfv(\bfr)$, is not necessarily spin-conservative. In other words, there may be a non-negligible effective spin-orbit coupling taking part in the hybridization. Such a scenario is reasonable in conjunction with chiral molecules and metals with strong spin-orbit coupling.

Since the self-energy comprises several parts, the conclusions drawn from the expressions in Eq. \eqref{eq-s} have to be reformulated. The induced spin-moment in the metal does not have to originate from the molecule \emph{per se}, it may actually arise from spin-dependent hybridization factors $\bfv_\bfk$. However, since the presence of the molecule in absence of the hybridization does not have any influence on the metallic properties, henceforth, the composite combination of the molecule and hybridization as a whole is referred to as the molecule.

\subsection{Molecular spin-polarization}
The first component in the expansion of $\av{\bfs(\bfr)}$ is interesting from the perspective that a local spin-polarization should be present in the vicinity of the molecule, given that the molecule is spin-polarized. However, as far as a self-consistent build up of a spin-polarization in both the substrate and the molecule is concerned, neither of the components $\av{\bfs(\bfr)}_\text{mol}$ and $\av{\bfs(\bfr)}_\text{sub}$ provide a deeper insight to this issue. This build up has to be considered in terms of the mixed component $\av{\bfs(\bfr)}_\text{mix}$.

In this section, the system is assumed to have reached a stationary state which allows for Fourier transforming out the time-variable. Then, the unperturbed Green function for the electrons in the metal is given by
\begin{align}
\bfg_\bfk(z)=&
	\frac{[z-\dote{0}(\bfk)]\sigma_0+\bfepsilon_1(\bfk)\cdot\bfsigma}
		{[z-\dote{0}(\bfk)]^2-\bfepsilon_1^2(\bfk)}
	.
\label{eq-gk}
\end{align}
Setting $\bfepsilon_1(\bfk)=\alpha\bfk$, the corresponding real space Green function $\bfg^r(\bfr;\omega)=\int\bfg_\bfk^r(\omega)e^{i\bfk\cdot\bfr}d\bfk/(2\pi)^d=g_0^r(\bfr;\omega)\sigma_0+\bfg_1^r(\bfr;\omega)\cdot\bfsigma$ is analytically calculable and for $d=3$ given by
\begin{subequations}
\label{eq-gr}
\begin{align}
g_0^r(\bfr;\omega)=&
	(-i)\frac{N_0}{4\pi}\sum_{s=\pm}\frac{\kappa_s^2}{\kappa}
	h_0^{(1)}(\kappa_s r)
	,
\label{eq-g0r}
\\
\bfg_1^r(\bfr;\omega)=&
	-\frac{N_0}{4\pi}\sum_{s=\pm}s\frac{\kappa_s^2}{\kappa}
	h_1^{(1)}(\kappa_s r)	
	\hat\bfr
		.
\label{eq-g1r}
\end{align}
\end{subequations}
In these expressions, $\kappa_s=\sqrt{2N_0(\omega+\dote{F}-\alpha^2N_0/2)}+s\alpha N_0$ with $N_0=m_e/\hslash^2$, whereas $h_\nu^{(n)}(\omega)$, $n=1,2$, is the $\nu$th spherical Hankel function of the $n$th kind. For later purposes, it should be noticed that $g_0$ ($\bfg_1$) is even (odd) under inversion symmetry operation, that is $g_0(-\bfr)=g_0(\bfr)$ and $\bfg_1(-\bfr)=-\bfg_1(\bfr)$, which also means that $\bfg_1(\bfr=0)=0.$

Then, the induced spin-polarization $\av{\bfs(\bfr)}_\text{mol}$ near a molecule located at $\bfr_0=0$ becomes
\begin{align}
\av{\bfs(\bfr)}_\text{mol}=&
	i\frac{N_0^2}{16\pi^2}
	\int
		\left|
			\sum_{s=\pm}
			\frac{\kappa_s^2}{\kappa}
			h_0^{(1)}(r\kappa_s)
		\right|^2
		\bfSigma_1^<(\omega)
	\frac{d\omega}{2\pi}
\nonumber\\\approx&
	i\frac{N_0^2}{4\pi^2}
	\int
		\kappa
		\Bigl|
			h_0^{(1)}(r\kappa)
		\Bigr|^2
		\bfSigma_1^<(\omega)
	\frac{d\omega}{2\pi}
	,
\label{eq-smolr}
\end{align}
where the approximated value in the last line is obtained under the condition that $\alpha^2N_0/2\ll\dote{F}$, which is reasonable for Au \cite{PhysRevB.83.165401,AdvMater.9.202201102}. Hence, a molecular spin-polarization transfers to and spreads isotropically in the substrate as a spin density wave. Crucially, no localized magnetic moment emerges in the substrate, merely a weak proximity induced evanescent spin-polarization decaying as $1/r^2$ in the substrate.

The induced spin-polarization originating from the substrate, $\av{\bfs(\bfr)}_\text{sub}$, requires a bit more care in the calculation. First it can be noticed that the last term in Eq. \eqref{eq-ssub} vanishes using the Green function in Eq. \eqref{eq-gk}, hence, the spin-moment is purely governed by the preexisting spin-texture captured in $\bfG_{\bfk\bfk'}^{1,r/a}$. Explicitly, this spin-polarization can be expressed as 
\begin{align}
\av{\bfs(\bfr)}_\text{sub}=&
	\biggl(\frac{N_0}{4\pi}\biggr)^2
	\im
	\int
		\biggl(
			\frac{\kappa_+^2}{\kappa}
			h_0^{(1)}(\kappa_+r)
			+
			\frac{\kappa_-^2}{\kappa}
			h_0^{(1)}(\kappa_-r)
		\biggr)
\nonumber\\&\times
		\biggl(
			\frac{\kappa_+^2}{\kappa}
			h_1^{(2)}(\kappa_+r)
			-
			\frac{\kappa_-^2}{\kappa}
			h_1^{(2)}(\kappa_-r)
		\biggr)
		\Sigma_0^<(\omega)
	\frac{d\omega}{2\pi}
	\hat\bfr
	.
\label{eq-ssubr}
\end{align}

It is worth emphasizing that this moment arises solely from the presence of the non-magnetic component of the adsorbant due to the spin-orbit coupling in the substrate. However, since it is emanating radially as $1/r^2$ from the location of the adsorbant it cannot sustain a fixed moment in any specific direction. The conclusion from analyzing this contribution is that, the presence of the molecule leads to the emergence of a spin-density wave and not a magnetic moment.

Finally, the spin-moment $\av{\bfs(\bfr)}_\text{mix}$ arising from the mixture of the molecular spin-moment and metallic spin-texture is given by
\begin{widetext}
\begin{align}
\av{\bfs(\bfr)}_\text{mix}=&
	i
	\biggl(\frac{N_0}{4\pi}\biggr)^2
	\sum_s
		\frac{\kappa_s^2}{\kappa}
		\Biggl(
			h_1^{(1)}(\kappa_sr)
			\biggl(
				2\bfSigma^{1,<}(\omega)
				\cdot\hat\bfr
				\hat\bfr
				-
				\bfSigma^{1,<}(\omega)
			\biggr)
			+
			s
			2
			h_0^{(1)}(\kappa_sr)
			\hat\bfr
			\times
			\bfSigma^{1,<}(\omega)
		\Biggr)
		\biggl(
			h_1^{(2)}(\kappa_sr)
			-
			\frac{\kappa_{\bar{s}}}{\kappa_s}
			h_1^{(2)}(\kappa_{\bar{s}}r)
		\biggr)
	.
\label{eq-smixr}
\end{align}
\end{widetext}
This moment acquires evanescent spin-density wave components governed by the spin-orbit coupling ($\propto\hat\bfr$), an induced spin moment along the effective moment of the molecular spin ($\propto\bfSigma^{1,<}$), as well as a third component which is perpendicular to both former components ($\propto\hat\bfr\times\bfSigma^{1,<}$).

In the context of an emergent local spin moment that may establish a spinterface between the molecule and the substrate, this mixed component is the most interesting. The idea behind the spinterface model is that a small Biot-Savart field breaks the spin-degeneracy of the conduction electrons in the (Au) substrate. By a strong spin-orbit coupling in the noble metal, the broken symmetry eventually accumulates to a strong spin moment aligned either parallel or anti-parallel with the longitudinal global spin direction defined by the set-up.

From this point of view, the contributions proportional to $\bfSigma^{1,<}$ and $\hat\bfr\times\bfSigma^{1,<}$ are more interesting to analyze, since the induce field can be associated with $\bfSigma^{1,<}$. However, this field is in the construction of the spinterface assumed to be weak. Since there are no distinctive anisotropies in the substrate that can pick up and amplify this small field, it is clear that the analysis can only render a negative result for the spinterface to serve as an explanation of the chirality induced spin selectivity effect.

\subsection{Effect of electron correlations}
\label{ssec-correlations}
A deeper insight to the properties of the electrons in the substrate and its capacity to sustain a large induced spin moment can be considered by introducing electron-electron interactions of type $U(\bfr-\bfr')n(\bfr)n(\bfr')$, where $U(\bfr)$ denotes the Coulomb repulsion energy and where $n(\bfr)=\psi^\dagger(\bfr)\psi(\bfr)$ defines the electronic occupation number at $\bfr$. Given that the Coulomb repulsion only depends on the distance $\bfr-\bfr'$ leads to that the corresponding model in reciprocal space assumes the form of the homogenous electron gas, expressed in the Hamiltonian
\begin{align}
\Hamil_\text{sub}=&
	\sum_\bfk
	\Bigl(
		\psi_\bfk^\dagger\bfepsilon_\bfk\psi_\bfk
		+
		U_\bfk\rho_\bfk\rho_{\bar\bfk}
	\Bigr)
	.
\end{align}
In this expression, $\rho_\bfk=\sum_\bfq\psi_{\bfk+\bfq}^\dagger\psi_\bfq$ denotes the electron density operator, whereas $U_\bfk=\int U(\bfr)e^{i\bfk\cdot\bfr}d\bfr$.

This model can diagonolized by introducing the projection operator $\X{pq}{}=\ket{p}\bra{q}$, where $p,q$ are state labels. First, setting $\psi_\bfk=(\sigma^0\ \sigma^z)\X{}{\bfk}$, where $\X{}{\bfk}=(\X{0\up}{\bfk},\ \X{0\down}{\bfk},\ \X{\down2}{\bfk},\ \X{\up2}{\bfk})^t$, captures single electron transitions between the available states for each $\bfk$. Accordingly, the homogeneous electron gas is re-expressed as
\begin{align}
\Hamil_\text{sub}=&
	\sum_\bfk
	\left(
		\sum_pE_p(\bfk)\X{pp}{\bfk}
		+
		\dote{\bfk-}\X{\up\down}{\bfk}
		+
		\dote{\bfk+}\X{\down\up}{\bfk}
	\right)
	,
\end{align}
where the label $p\in\{0,\up,\down,2\}$ runs over the empty state, and the singly and doubly occupied states with corresponding energies $E_0(\bfk)=0$, $E_\up(\bfk)=\dote{0}(\bfk)+\dote{z\bfk}+U_\bfk$, $E_\up(\bfk)=\dote{0}(\bfk)-\dote{z\bfk}+U_\bfk$, and , $E_\up(\bfk)=2\dote{0}(\bfk)+4U_\bfk$. Furthermore, the spin-flip energies $\dote{\bfk\pm}=\dote{\bfk x}\pm i\dote{\bfk y}$ represent the transversal part of the vector $\bfepsilon_1(\bfk)=(\dote{\bfk x},\dote{\bfk y},\dote{\bfk z})$. Finally, by the unitary transformation
\begin{align}
	\begin{pmatrix}
		\X{\up\up}{\bfk} \\
		\X{\down\down}{\bfk}
	\end{pmatrix}
	=&
	\begin{pmatrix}
		\cos\theta_\bfk & -e^{-i\phi_\bfk}\sin\theta_\bfk \\
		e^{i\phi_\bfk}\sin\theta_\bfk & \cos\theta_\bfk
	\end{pmatrix}
	\begin{pmatrix}
		\X{++}{\bfk} \\
		\X{--}{\bfk}
	\end{pmatrix}
	,
\end{align}
the diagonalization is complete, giving
\begin{align}
\Hamil_\text{sub}=&
	\sum_{\stackrel{\scriptstyle \bfk}{p=0,\pm,2}}
		E_p(\bfk)\X{pp}{\bfk}
	,
\end{align}
with $E_\pm(\bfk)=\dote{0}(\bfk)+U_\bfk\pm\dote{1}(\bfk)$, $\dote{1}(\bfk)=|\bfepsilon_1(\bfk)|$.

This diagonal form allows for analytical calculations, where the partition function per $\bfk$ is given by
\begin{align}
Z_0(\bfk)=&
	1
	+
	\Bigl(
		2\cosh\beta\dote{1}(\bfk)
		+
		e^{-\beta(\dote{0}(\bfk)+3U_\bfk)}
	\Bigr)
	e^{-\beta(\dote{0}(\bfk)+U_\bfk)}
	,
\end{align}
and the occupation numbers
\begin{subequations}
\begin{align}
N_0(\bfk)=&\frac{1}{Z_0(\bfk)}
	,\
N_2(\bfk)=
	\frac{e^{-2\beta(\dote{0}(\bfk)+2U_\bfk)}}{Z_0(\bfk)}
\\
N_\up(\bfk)=&
	N_1(\bfk)-N_z(\bfk)
	,\
N_\down(\bfk)=
	N_1(\bfk)+N_z(\bfk)
	,
\\
N_1(\bfk)=&
	\frac{e^{-\beta(\dote{0}(\bfk)+U_\bfk)}}{Z_0(\bfk)}
	\cosh\beta\dote{1}(\bfk)
	,
\\
N_z(\bfk)=&
	N_1(\bfk)
	\frac{\dote{\bfk z}}{\dote{1}(\bfk)}
	\tanh\beta\dote{1}(\bfk)
	.
\end{align}
\end{subequations}
Consequently, the spin-polarization is determined from the difference
\begin{align}
N_\up(\bfk)-N_\down(\bfk)=&
	-2
	\frac{\dote{\bfk z}}{\dote{1}(\bfk)}
	\frac{e^{-\beta(\dote{0}(\bfk)+U_\bfk)}}{Z_0(\bfk)}
	\sinh\beta\dote{1}(\bfk)
	.
\end{align}

First it can be noticed that the spin-polarization varies non-monotonically with the Coulomb interaction $U_\bfk$, with a local maximum value around $\dote{0}(\bfk)+U_\bfk=0$. For $\dote{0}(\bfk)+U_\bfk<0$, the occupation $N_2(\bfk)$ of the two-particle state $\ket{\bfk,2}$ quenches the spin-polarization. In the strongly correlated limit $\dote{0}(\bfk)+U_\bfk\gg0$, on the other hand, the spin-polarization is exponentially suppressed by the interactions.

Second, considering the presence of a spin-orbit coupling $\alpha\bfk$ in the substrate and an external magnetic field $\bfB=B\hat{\bf z}$, the vector $\bfepsilon_1(\bfk)=\alpha\bfk-g\mu_BB\hat{\bf z}$ has the modulus $\dote{1}(\bfk)=\sqrt{\alpha^2k^2-2\alpha k_zg\mu_BB+(g\mu_BB)^2}$. For magnetic field strengths $g\mu_BB\ll\alpha k$, then $\dote{1}(\bfk)\approx\alpha k$, which suggests that the spin-polarization grows linearly with $B$. Under the opposite conditions, $g\mu_BB\gg\alpha k$, the induced spin-polarization 

By contrast, under the same conditions, the spin-polarization depends only weakly on the strength of the spin-orbit coupling, since the ratio $\dote{\bfk z}/\dote{1}(\bfk)\approx(\alpha k_z-g\mu_BB)/\alpha k$ whereas $e^{-\beta(\dote{0}(\bfk)+U_\bfk)}\sinh\beta\dote{1}(\bfk)/Z_0(\bfk)\approx 1/2$ for large $\dote{1}(\bfk)$. In general, the latter ratio is less than unity such that the induced spin-polarization is $g\mu_BB$ at most.

Hence, while a weak magnetic field induces a small spin-polarization in the substrate, which only depends linearly on the magnetic field strength at most, a strong spin-orbit coupling cannot amplify this induced spin-polarization. In addition, since the induced spin-polarization decays as $\sinh\beta\dote{1}(\bfk)$ with increasing temperature, the magnetic field strength $B$ would have to be in the order of several hundreds of Tesla in order to stabilize a significant spin-polarization at room temperature.

\subsection{Spin dynamics}
The previous discussion is based on time-independent, or stationary, conditions which can only reflect the long-time properties of the interface. Any type transient of dynamical behavior of the spin-density which is omitted has to be considered in a study of the time-dependent properties, which is the purpose of this section.

The dynamics of the local spin moment $\av{\bfs(\bfr,t)}$ is here captured through the study of $\av{\bfs_{\bfk\bfk'}(t)}$ which is enabled by the relation $\av{\bfs(\bfr,t)}=\int\av{\bfs_{\bfk\bfk'}(t)}e^{-i(\bfk-\bfk')\cdot\bfr}d\bfk d\bfk'/\Omega^2$, where $\Omega$ denotes the volume of integration. Furthermore, the expectation values of the electron spin $\av{\bfs_{\bfk\bfk'}(t)}$ can be studied in terms of the single electron Green function $\bfG_{\bfk\bfk'}(t,t')$ thanks to the relations $\av{\bfs_{\bfk'\bfk}}(t)=(-i)\lim_{t'\rightarrow t^+}{\rm sp}\bfsigma\bfG^<_{\bfk\bfk'}(t,t')/2$ and $i\dt\av{\bfs_{\bfk'\bfk}}(t)=(-i)\lim_{t'\rightarrow t^+}{\rm sp}\bfsigma(i\dt+i\partial_{t'})\bfG^<_{\bfk\bfk'}(t,t')/2$.

It it straight forward to derive
\begin{widetext}
\begin{align}
i\dt
	\bfG^<_{\bfk\bfk'}(t)=&
	\bfepsilon_\bfk\bfG^<_{\bfk\bfk'}(t)
	-
	\bfG^<_{\bfk\bfk'}(t)\bfepsilon_{\bfk'}
\label{eq-dtGlesser}\\&
	+
	\sum_\bfp
	\int_{-\infty}^t
	\biggl(
		\bfv_\bfk\bfG_1^>(t,\tau)\bfv^\dagger_\bfp\bfG^<_{\bfp\bfk'}(\tau,t)
		-\bfv_\bfk\bfG_1^<(t,\tau)\bfv^\dagger_\bfp\bfG^>_{\bfp\bfk'}(\tau,t)
		-
		\bfG^>_{\bfk\bfp}(t,\tau)\bfv_\bfp\bfG^<_1(\tau,t)\bfv^\dagger_{\bfk'}
		+
		\bfG^<_{\bfk\bfp}(t,\tau)\bfv_\bfp\bfG^>_1(\tau,t)\bfv^\dagger_{\bfk'}
	\biggr)
	d\tau
	.
\nonumber
\end{align}
Here, taking the trace over spin 1/2 space of the first two terms on the right hand side gives the contribution
\begin{align}
(-i)\frac{1}{2}
	{\rm sp}\bfsigma
	\biggl(
		\bfepsilon_\bfk\bfG^<_{\bfk\bfk'}-\bfG^<_{\bfk\bfk'}\bfepsilon_{\bfk'}
	\biggr)
	=&
		\dote{\bfk\bfk'}\av{\bfs_{\bfk'\bfk}}
		+
		i\bfD_{\bfk\bfk'}\times\av{\bfs_{\bfk'\bfk}}
		+
		\bfC_{\bfk\bfk'}\av{\psi^\dagger_{\bfk'}\psi_\bfk}
	.
\end{align}
where $\dote{\bfk\bfk'}=\dote{0}(\bfk)-\dote{0}(\bfk')$, $\bfD_{\bfk\bfk'}=\bfepsilon_1(\bfk)+\bfepsilon_1(\bfk')$, and $\bfC_{\bfk\bfk'}=[\bfepsilon_1(\bfk)-\bfepsilon_1(\bfk')]/2$.
Moreover, making use of the time-independent conditions in the set-up, the last term in Eq. (\ref{eq-dtGlesser}) can be represented by
\begin{align}
\bfj_{\bfk\bfk'}=&
	\sum_\bfp
	\int
	\left(
		\frac{\bfG^>_{\bfk\bfp}(\omega)\bfv_\bfp\bfG^<_1(\omega')-\bfG^<_{\bfk\bfp}(\omega)\bfv_\bfp\bfG^>_1(\omega')}{\omega-\omega'-i\delta}
		\bfv^\dagger_{\bfk'}
		-
		\bfv_\bfk
		\frac{\bfG^>_1(\omega)\bfv^\dagger_\bfp\bfG^<_{\bfp\bfk'}(\omega')-\bfG^<_1(\omega)\bfv^\dagger_\bfp\bfG^>_{\bfp\bfk'}(\omega')}{\omega-\omega'-i\delta}
	\right)
	\frac{d\omega}{2\pi}
	\frac{d\omega'}{2\pi}
\nonumber\\=&
	j_0(\bfk\bfk')\sigma^0+\bfj_1(\bfk\bfk')\cdot\bfsigma
	,
\end{align}
\end{widetext}
where $j_0$ and $\bfj_1$ represent the charge and spin currents, respectively, flowing across the interface between the substrate and molecule. For the dynamics of the electron spin, only the spin current contributes, since ${\rm sp}\bfsigma\bfj_{\bfk\bfk'}=2\bfj_1(\bfk\bfk')$, such that the rate of change for the electron spin $\av{\bfs_{\bfk'\bfk}}$ can be written
\begin{align}
\Bigl(
	i\dt-\dote{\bfk\bfk'}
	\Bigr)
	\av{\bfs_{\bfk'\bfk}}=&
	i\bfD_{\bfk\bfk'}\times\av{\bfs_{\bfk'\bfk}}
\nonumber\\&
	+
	\bfC_{\bfk\bfk'}\av{\psi^\dagger_{\bfk'}\psi_\bfk}
	+
	i\bfj_1(\bfk\bfk')
	.
\label{eq-skk}
\end{align}

It is clear that the electron charge $\av{\psi^\dagger_{\bfk'}\psi_\bfk}=(-i){\rm sp}\bfG^<_{\bfk\bfk'}$ has to be considered simultaneously with the electron spin. This can also be obtained from Eq. (\ref{eq-dtGlesser}), giving
\begin{align}
\Bigl(
	i\dt-\dote{\bfk\bfk'}
	\Bigr)
	\av{\psi^\dagger_{\bfk'}\psi_\bfk}=&
	4\bfC_{\bfk\bfk'}\cdot\av{\bfs_{\bfk'\bfk}}
	+
	i2j_0(\bfk\bfk')
	,
\label{eq-nkk}
\end{align}
where only the charge current contributes, as expected.

The first question to be addressed is the influence of the spin-orbit component in the electron energy $\bfepsilon_\bfk$, and finding a necessary condition for generating a finite spin selectivity in the current flow through the molecule. Hence, the intrinsic properties of the molecule remain constant under varying magnetic conditions of the metal. Then, in the equations for the electron charge and spin above, the contributions $i2j_0(\bfk\bfk')$ and $i\bfj_1(\bfk\bfk')$ will be treated as external forces.

By first omitting these forces, the homogenous equations can be written on the form
\begin{align}
\Bigl(
	i\dt-\mathbb{A}_{\bfk\bfk'}
	\Bigr)
	\begin{pmatrix}
	\av{\psi^\dagger_{\bfk'}\psi_\bfk} \\
	\av{\bfs_{\bfk'\bfk}}
	\end{pmatrix}
	=&
	0
	,
\end{align}
where the matrix
\begin{align}
\mathbb{A}_{\bfk\bfk'}=&
	\begin{pmatrix}
		\dote{\bfk\bfk'} & 4\bfC_{\bfk\bfk'} \\
		\bfC_{\bfk\bfk'}^t & \bfE_{\bfk\bfk'}
	\end{pmatrix}
	,
\end{align}
in which
\begin{align}
\bfE_{\bfk\bfk'}=&
	\Bigl(
		\delta_{ij}\dote{\bfk\bfk'}
		-
		i\dote{ijk}D_{\bfk\bfk'}^k
	\Bigr)
	\hat{\bf i}\hat{\bf j}
	,
\end{align}
is a $3\times3$-matrix.
%
Hence, in the homogenous system, the charge and spin densities are coupled only whenever $\bfepsilon_1(\bfk)\neq0$, as expected. Under conditions such that $\bfepsilon_1(\bfk)=0$, the matrix $\mathbb{A}_{\bfk\bfk'}=\dote{\bfk\bfk'}\mathbb{I}$ is diagonal and proportional to the unit matrix $\mathbb{I}$, which leads to that both densities $\av{\psi^\dagger_{\bfk'}\psi_\bfk}$ and $\av{\bfs_{\bfk'\bfk}}$ are constants of motion; individually conserved quantities and good quantum numbers.

The eigenvalues of the matrix $\mathbb{A}_{\bfk\bfk'}$ are analytically accessible and given by
\begin{align}
\lambda_{ss'}(\bfk,\bfk')=&
	\dote{0}(\bfk)-\dote{0}(\bfk')
	+
	s
	\sqrt{
		\Bigl(
			\dote{1}(\bfk)
			+
			s'\dote{1}(\bfk')
		\Bigr)^2
	}
	,
\end{align}
where $\dote{1}(\bfk)=\sqrt{\bfepsilon_1(\bfk)\cdot\bfepsilon_1(\bfk)}$ and where $s,s'=\pm1$. Hence, in presence of the inhomogeneous components, that is, including the currents $j_0$ and $\bfj_1$, the solution to the system of equation given by Eqs. \eqref{eq-skk} and \eqref{eq-nkk} can be written as
\begin{align}
\begin{pmatrix}
	\av{\psi^\dagger_{\bfk'}\psi_\bfk} \\
	\av{\bfs_{\bfk'\bfk}}
\end{pmatrix}=&
	(-i)
	\sum_{ss'}
		\frac{1-e^{-i\lambda_{ss'}(\bfk,\bfk')(t-t_0)}}{\lambda_{ss'}(\bfk,\bfk')}
\nonumber\\&
		\times
		\bfv_{ss}(\bfk,\bfk')
		\begin{pmatrix}
			\av{\psi^\dagger_{\bfk'}\psi_\bfk}_0+i2j_0(\bfk\bfk')\\
			\av{\bfs_{\bfk'\bfk}}_0+i\bfj_1(\bfk\bfk')
		\end{pmatrix}
	,
\end{align}
where $\bfv_{ss}(\bfk,\bfk')$ denotes the eigenstates corresponding to the eigenvalues $\lambda_{ss}(\bfk,\bfk')$, and where $\av{\psi^\dagger_{\bfk'}\psi_\bfk}_0$ and $\av{\bfs_{\bfk'\bfk}}_0$ are given by the initial conditions at $t=t_0$. Here, it should, moreover, be noticed that the energies $\dote{0}$ and $\bfepsilon_1$ as well as the currents $j_0$ and $\bfj_1$ are assumed to be time-independent.

The eigenvalues $\lambda_{ss'}(\bfk,\bfk')$ are all real numbers. This implies that both densities $\av{\psi_{\bfk'}^\dagger\psi_\bfk}$ and $\av{\bfs_{\bfk'\bfk}}$ are merely oscillating functions of time, irrespective of whether the currents $j_0$ and $\bfj_1$ are non-vanishing or not. Hence, the charge and spin densities repeatedly vary between different configurations and never assume a stationary phase. The conclusion is, therefore, that an electron flux, spin-polarized or not, in to or out from the molecule does not lead to a stabilized spin density at the interface between the molecule and the substrate.

Despite this negative result for the stabilization of a spinterface, the conditions change dramatically should the eigenvalues comprise imaginary parts as well. Namely, imaginary parts of the eigenvalues represent irreversible processes, e.g., damping or driving forces. Under such conditions, the system will eventually assume a stationary phase, which may be the same as the initial. On the other hand, because of the spin-orbit coupling $\bfepsilon_1(\bfk)$, none of the eigenvalues correspond to a pure spin state which means that in whatever state the system eventually finds it stationary phase, there is no fixed spin-moment associated with this state.

It should be clear, that any non-linearity that might lead to a finite spin accumulation requires particle-particle interactions. Hence, studying the time-evolution of this process is beyond the capacity of the simple free electron model used for conduction electrons in the substrate. On the other hand, the discussion in Sec. \ref{ssec-correlations} strongly suggests that a Coulomb interaction of the type that is captured by the homogeneous electron gas does not sustain and stabilize a local magnetic moment, despite an initial small symmetry breaking field acting from external sources.


\section{Summary and conclusions}
The electronic and magnetic properties near and around the interface between a metallic substrate and a molecular adsorbant have been studied in three different and complementary ways, all pointing towards the same conclusion: A strong spin-orbit coupling in the substrate cannot stabilize and sustain a large magnetic moment irrespective of whether there is a symmetry breaking agent, e.g., small magnetic field, that would induced such moment. In spite of the fact that there have been a few reports of noble metal nanoparticles providing magnetic responses, even ferromagnetic such, it is highly questionable that, for instance, a Au layer decorated with a self-assembled mono-layer of chiral molecules would generate a strongly spin-polarized interface with these molecules.

It is clear that neither spin-orbit coupling nor Coulomb interactions by themselves may break the local symmetry and, hence, spawn a spin-polarized state. Even so, not even a concerted effect by these mechanisms is likely to generate the local magnetic moment required for the spinterface effect alluded to in Refs. \citenum{JACS.143.14235,JACS.146.32795,ACSNano.19.37484} as a prerequisite to explain the chirality induced spin selectivity effect. Furthermore, recent first principles calculations on (M/P)-heptahelicene adsorbed on Au surface do not display any signatures that there would emerge an interface spin-polarization \cite{2603.22725}.

While the presented material may not be sufficiently conclusive to entirely dismiss the spinterface model as a possible explanation for the chirality induced spin selectivity effect, it conclusively states that should this model be viable, its origin is far more profound and deeply lying than what is accessible using standard formalisms. Since the localized $d$-levels are both sufficiently far below the Fermi level to be essentially passivated in the context of both transport and photo-electron spectroscopy \cite{2603.22725}, any emergent spin-polarization has to be associated with itinerant $s$- and $p$-electrons. In this sense, the presented material applies generally to any metallic materials, including graphene, Cu, Ag, and Au, where bands associated with localized electrons can be omitted in the transport calculations.

Although the present study indicates that a spin-polarization in the adsorbant would induce a small spin-polarization in the substrate as well, it will not be become sufficiently large to be accounted for in a spinterface model. Thereto, the induced spin-polarization is not strongly localized and cannot be treated as a classical spin, which prevents the application of the Landau-Lifshitz-Gilbert equation. Resolving the issue of an induced strong interfacial magnetic moment will, most likely, have to be a question for first principles calculations.


\acknowledgements
The author thanks A. Bergman, Y. Dubi, O. Eriksson, P. Hedeg\aa rd, Y. Geerts, R. Naaman, L. Nordstr\"om, Y. Paltiel, and D. Waldeck for fruitful discussions on the matter presented in this article. Funding from Olle Engkvists Stiftelse is acknowledged.

\bibliography{CISSref}

\end{document}